\documentclass[final,3p,times,twocolumn]{elsarticle}

\usepackage{hyperref,siunitx}
\usepackage{lineno}
\usepackage{amsmath,amssymb}
\usepackage{caption}
\usepackage{subcaption}
\usepackage[version=4]{mhchem}

\DeclareSIUnit{\neq}{n_{eq}}
\DeclareSIUnit{\e}{e^{-}}
\DeclareSIUnit{\rad}{rad}

\modulolinenumbers[1]

\begin{document}
	
\begin{frontmatter}
	
	\title{DMAPS Monopix developments in large and small electrode designs}
	
	\author[1]{C. Bespin\corref{cor1}}%
    \ead{bespin@physik.uni-bonn.de}
    \author[2]{M. Barbero}
    \author[2]{P. Barrillon}
    \author[4]{I. Berdalovic}
    \author[2]{S. Bhat}
    \author[2]{P. Breugnon}
	\author[1]{I. Caicedo}
	\author[4]{R. Cardella}
	\author[2]{Z. Chen}
	\author[3]{Y. Degerli}
	\author[1]{J. Dingfelder}
    \author[4]{L. Flores Sanz de Acedo}
	\author[2]{S. Godiot}
	\author[3]{F. Guilloux}
	\author[1]{T. Hirono}
	\author[1]{T. Hemperek}
    \author[1]{F. H\"ugging}
	\author[1]{H. Kr\"uger}
	\author[4]{T. Kugathasan}
	\author[4]{C. Marin Tobon}
	\author[1]{K. Moustakas}
	\author[2]{P. Pangaud}
	\author[4]{H. Pernegger}
	\author[4]{F. Piro}
	\author[4]{P. Riedler}
	\author[2]{A. Rozanov}
	\author[1]{P. Rymaszewski}
	\author[3]{P. Schwemling}
	\author[4]{W. Snoeys}
	\author[3]{M. Vandenbroucke}
	\author[1]{T. Wang}
	\author[1]{N. Wermes}
	\author[1]{S. Zhang}
	
	\cortext[cor1]{Corresponding author}
	
	\address[1]{Universit\"at Bonn, Physikalisches Institut, Nu{\ss}allee 12, 53115 Bonn, Germany}
	\address[2]{Aix Marseille University, CNRS/IN2P3, CPPM, 163 Avenue de Luminy, 13009 Marseille, France}
	\address[3]{IRFU, CEA-Saclay, 91191 Gif-sur-Yvette, France}
	\address[4]{CERN, Espl. des Particules 1, 1211 Meyrin, Switzerland}
	
	\begin{abstract}
		LF-Monopix1 and TJ-Monopix1 are depleted monolithic active pixel sensors (DMAPS) in~\SI{150}{nm} LFoundry and~\SI{180}{nm} TowerJazz CMOS technologies respectively.
		They are designed for usage in high-rate and high-radiation environments such as the ATLAS Inner Tracker at the High-Luminosity Large Hadron Collider (HL-LHC).
		Both chips are read out using a column-drain readout architecture.
		LF-Monopix1 follows a design with large charge collection electrode where readout electronics are placed inside. 
		Generally, this offers a homogeneous electrical field in the sensor and short drift distances.
		TJ-Monopix1 employs a small charge collection electrode with readout electronics separated from the electrode and an additional n-type implant to achieve full depletion of the sensitive volume.
		This approach offers a low sensor capacitance and therefore low noise and is typically implemented with small pixel size.
		Both detectors have been characterized before and after irradiation using lab tests and particle beams.
	\end{abstract}


\end{frontmatter}

% \linenumbers

\section{Introduction}
Monolithic pixel sensors designed in commercial CMOS technologies were first proposed in the 1990s for charged particle tracking.
They offer low material budget, low cost and easy module assembly compared to well established hybrid pixel detectors~\cite{turchetta2001677}.
However, monolithic active pixel sensors, for instance the ALPIDE chip for the ALICE experiment at LHC~\cite{mager2016434} are not suitable for high-radiation and high-rate environments as their charge collection mechanism is mainly by diffusion.
In order to cope with conditions such as those expected at the future HL-LHC, charge collection has to be done by drift in a depleted sensor volume, leading to depleted monolithic active pixel sensors (DMAPS).
Large-scale prototype chips employing high resistivity substrates and high bias voltages have been designed in various CMOS technologies with integrated fast readout electronics on the sensor substrate~\cite{lf_paper_toko, kiehn2019, Cardella_2019}.

Two of these devices, LF-Monopix1 and TJ-Monopix1, fabricated in \SI{150}{nm} LFoundry and~\SI{180}{nm} TowerJazz technology, respectively, have been characterized with respect to the requirements of the ATLAS ITk outer pixel layer.
Detectors in this environment have to withstand up to~\SI{e15}{\neq\per\centi\meter\squared} non-ionizing energy loss (NIEL) and~\SI{50}{\mega\rad} total ionizing dose (TID) radiation damage while maintaining timing requirements of~\SI{25}{\nano\second}.

\section{Design}
For a planar arrangement the depletion depth $d$ of the sensor volume depends on the resistivity $\rho$ of the material and the applied bias voltage $V$:
\begin{align*}
    d\propto\sqrt{\rho V}    
\end{align*}
Thus, to achieve full depletion of the sensing volume the design of DMAPS employs both highly resistive materials as well as high voltage features offered by the foundries.
In the following, the different sensor approaches of the reported pixel sensors are described.

\subsection{Large collection electrode in LFoundry technology}
LF-Monopix1 is designed in~\SI{150}{\nano\metre} CMOS technology and uses a so-called large collection electrode that encapsulates the pixel electronics.
The substrate material has a resistivity larger than~\SI{2}{\kilo\ohm\centi\meter} and can withstand bias voltages larger than~\SI{200}{\volt} while offering thicknesses down to~\SI{100}{\micro\meter}.
Due to the large n-type collection electrode, the sensing volume can be fully and homogeneously depleted while keeping drift paths in the volume short, and therefore creating a radiation-hard design.
This comes at a cost of the comparably large sensor capacitance of LF-Monopix1 of an estimated~${\sim\SI{400}{\femto\farad}}$ based on measurements of test structures~\cite{loepke} and a power consumption of $\sim$~\SI{36}{\micro\watt\per pixel} for a~\SI{250x50}{\micro\meter} pixel with full digital logic inside.
\begin{figure}[htb]
    \centering
    \includegraphics[width=0.9\columnwidth]{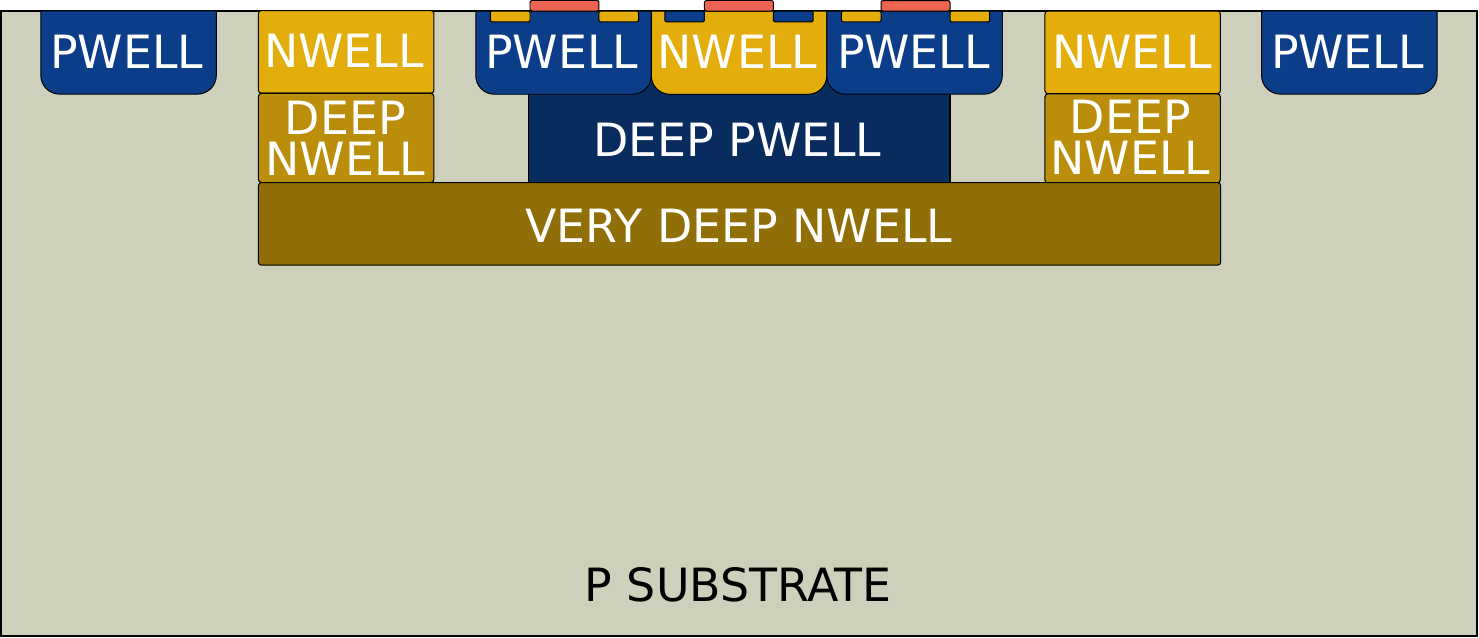}
    \caption{Schematic cross section of the LFoundry~\SI{150}{\nano\meter} process. Charge collection is done by an n-type implantation that encapsulates the readout electronics.}
    \label{fig:crosssection_lf}
\end{figure}
As depicted in figure~\ref{fig:crosssection_lf}, the in-pixel electronics make use of the nested well structure offered by the foundry.
In particular, the deep p-well implant allows for full CMOS logic, isolating the n-wells of PMOS transistors from the charge collection node.

\subsection{Small collection electrode in TowerJazz technology}
\begin{figure}[htb]
    \centering
    \begin{subfigure}{0.9\columnwidth}
    \centering
        \includegraphics[width=\linewidth]{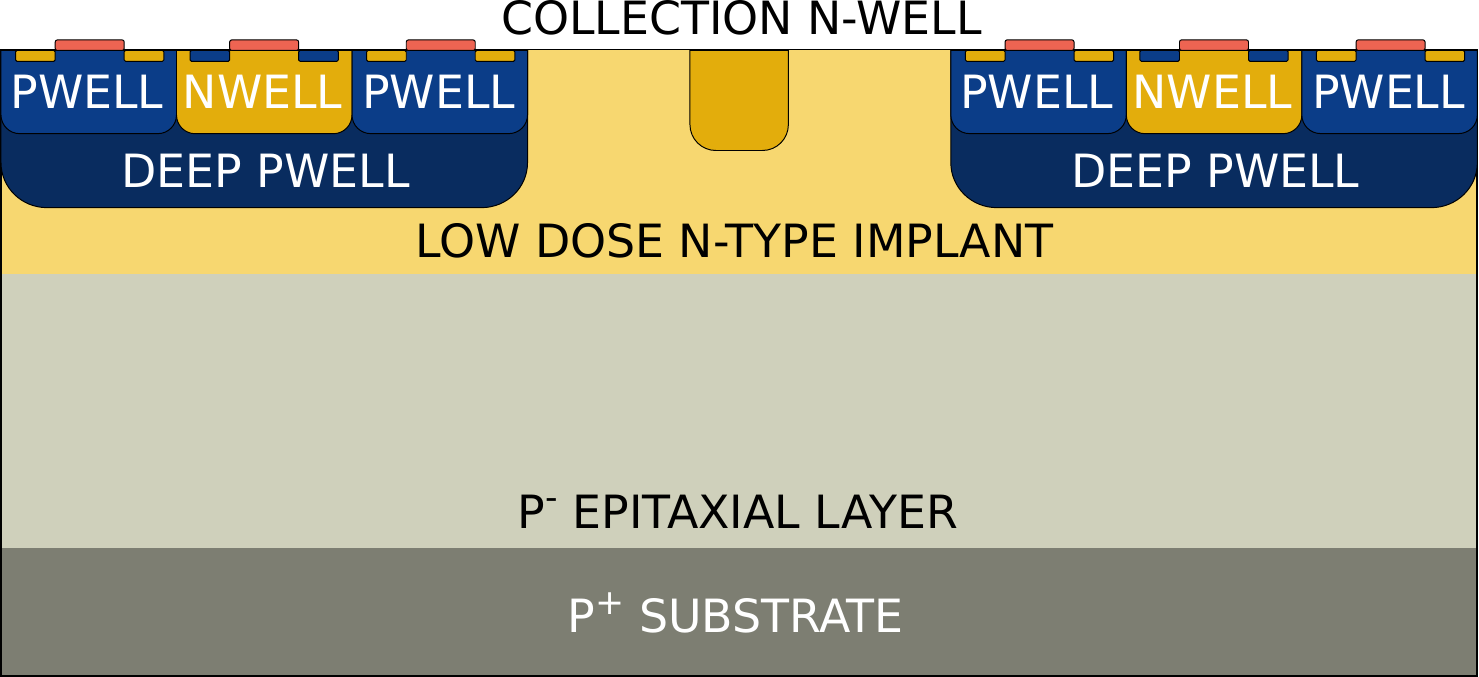}
        \caption{Sensor design with full deep p-well coverage.}
        \label{fig:crosssection_tj_fdpw}
    \end{subfigure}
    \begin{subfigure}{0.9\columnwidth}
    \centering
        \includegraphics[width=\linewidth]{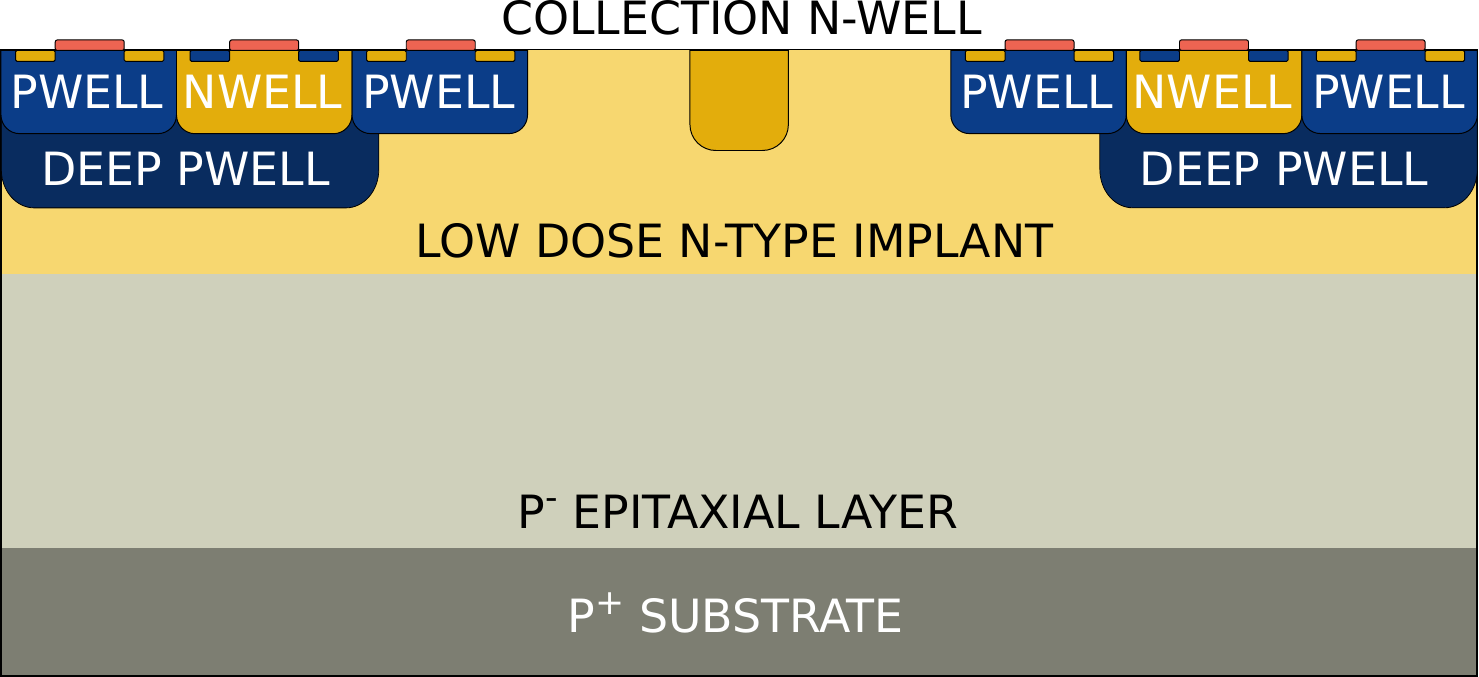}
        \caption{Sensor design with reduced deep p-well coverage.}
        \label{fig:crosssection_tj_rdpw}
    \end{subfigure}
    \caption{Schematic cross section of the TowerJazz~\SI{180}{\nano\meter} process. Charge collection is done by an n-type implantation which is separated from the readout electronics. CMOS transistors are shielded by a deep p-type implant to prevent the existence of a competing collection electrode from the n-wells of the PMOS transistors.}
    \label{fig:crosssection_tj}
\end{figure}

TJ-Monopix1 is designed in~\SI{180}{\nano\meter} TowerJazz technology and makes use of a small collection electrode which is separated from the pixel electronics.
It is based on the ALPIDE chip~\cite{mager2016434} designed for the ALICE experiment at LHC.
Figure~\ref{fig:crosssection_tj} shows a schematic cross section of this design, where charges are created in a~\SI{25}{\micro\meter} thick p-type epitaxial layer with a resistivity of more than~\SI{1}{\kilo\ohm\centi\meter}.

This approach offers a low sensor capacitance of~${\sim\SI{4}{\femto\farad}}$ allowing for a low power analog front-end design of \SI{1}{\micro\watt\per pixel}~\cite{moustakas2019604}, but is typically not as radiation hard as a large collection electrode design due to longer drift distances and inhomogeneous drifting fields.
In order to achieve full depletion and radiation hardness an additional low dose n-type layer has been implanted that extends the depletion region under the electronics on the pixel edge~\cite{snoeys201790}.
A deep p-well shields the n-well of PMOS transistors and prevent the existence of a competing collection electrode and allows for full CMOS logic of the electronics.
There are two variants of this deep p-well implant integrated in the chip: one that fully covers the in-pixel electronics and one that covers them only partially, which changes the shape of the electrical field in the sensor close to the charge collection electrode.
Results presented in this paper are obtained using pixels with reduced deep p-well coverage.

\begin{figure}[htb]
    \centering
    \begin{subfigure}{0.9\columnwidth}
    \centering
        \includegraphics[width=.9\linewidth]{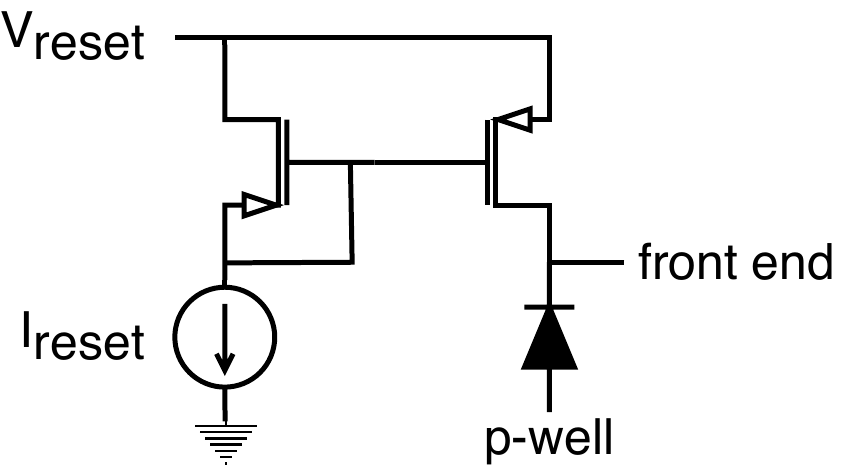}
        \caption{Front-end design with PMOS reset circuit and DC coupling.}
        \label{fig:pmos_reset}
    \end{subfigure}
    \begin{subfigure}{0.9\columnwidth}
    \centering
        \includegraphics[width=.8\linewidth]{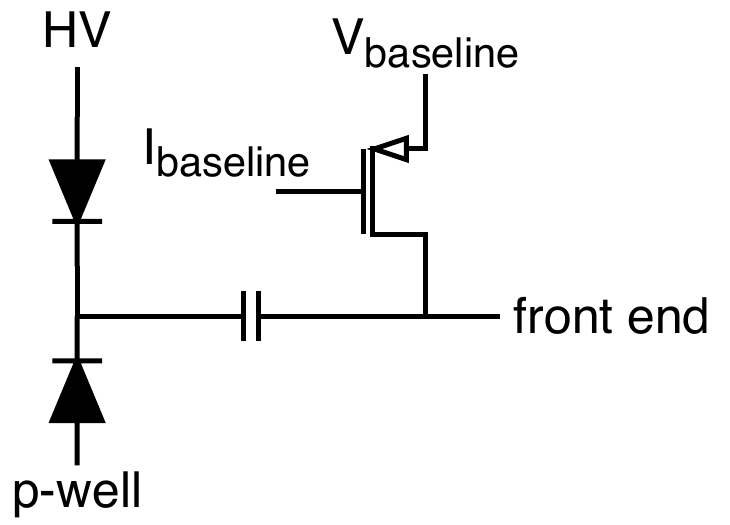}
        \caption{Front-end design with additional high voltage supply and AC coupling}
        \label{fig:hv_diode}
    \end{subfigure}
    \caption{Schematics of the pixel reset electronics and AC and DC coupling options for TJ-Monopix1.}
    \label{fig:reset_schematics}
\end{figure}
Three of the four available pixel flavors use slightly varying versions of a PMOS reset circuit that is DC coupled to the front-end.
One exemplary schematic is shown in figure~\ref{fig:pmos_reset}.
The fourth flavor utilizes an additional high voltage supply on a diode to reset the charge collection node and uses AC coupling to the front-end circuit~(\ref{fig:hv_diode}).

\subsection{Readout architecture}
Both LF-Monopix1 and TJ-Monopix1 make use of a column drain readout architecture, similar to the one used in the FE-I3 ATLAS pixel readout chip~\cite{PERIC2006178}.
The design is fully monolithic with all the dedicated readout electronics integrated into the pixel cell.
Charge created in the sensing volume and collected by the collection electrode is amplified and converted into a voltage pulse that is compared to an adjustable threshold.
The digital output of the discriminator defines a leading edge and trailing edge pulse determining the time of arrival and the time over threshold (ToT) -- the total length of the pulse.
Both are measured in units of clock cycles of~\SI{40}{\mega\hertz}.

\section{Measurements}
Various measurements on both detectors were performed before and after different types of irradiation.
NIEL damage studies were performed with chips irradiated with neutrons at the TRIGA Mark II Research Reactor of the Jožef Stefan Institute in Ljubljana~\cite{jsi} up to~\SI{e15}{\neq\per\centi\meter\squared}.
The TID tolerance has been tested with X-ray irradiated chips up to~\SI{100}{\mega\rad}.
Efficiency results were obtained using a~\SI{2.5}{\giga\electronvolt} beam at the External Beamline for Detector Tests at the Electron Stretcher Accelerator at Bonn University~\cite{Heurich:IPAC2016-THPOY002}.

\subsection{Results for LF-Monopix1}

LF-Monopix1 has been characterized in regard to the pixel electronics' performance such as gain and noise.
These measurements were conducted with untuned sensors cooled down to approximately~\SI{-25}{\degreeCelsius}.
Additional results are given on the hit detection efficiency measured during a test beam campaign.
Results for irradiated sensors have been obtained using the same chip as in~\cite{lf_paper_toko}.

\begin{figure}[htb]
    \centering
    \includegraphics[width=.9\columnwidth]{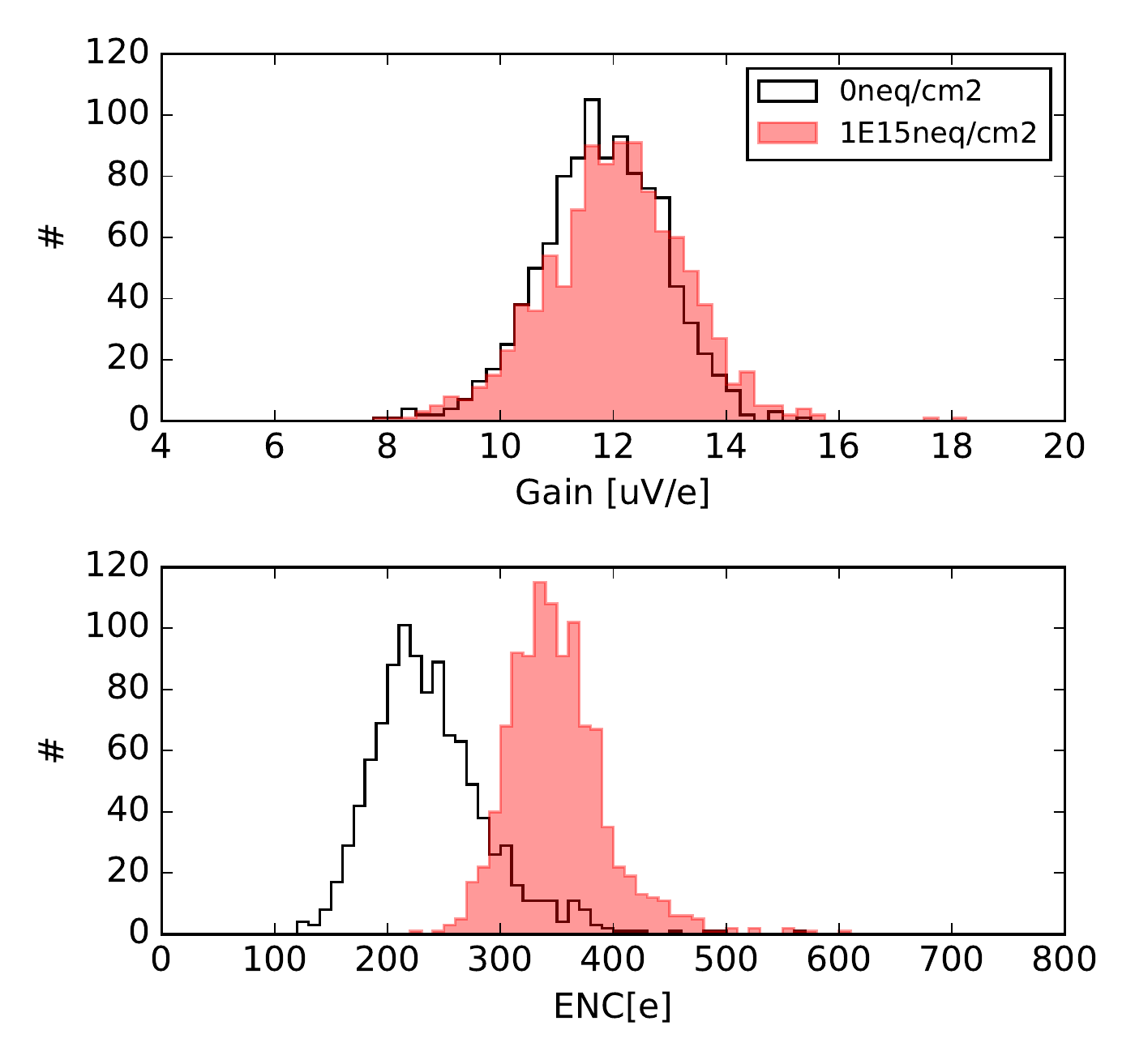}
    \caption{Gain and noise distribution of LF-Monopix1 before and after neutron irradiation to~\SI{e15}{\neq\per\square\centi\meter}.}
    \label{fig:lf_gain_noise_irrad}
\end{figure}

The gain of LF-Monopix1 has been measured as voltage response of the amplifier to a~\ce{^{55}Fe} signal.
Gain distributions before and after neutron irra\-diation to~\SI{e15}{\neq\per\square\centi\meter} are depicted in figure~\ref{fig:lf_gain_noise_irrad} (top) with no observable change due to radiation.
Injecting an external voltage pulse into each pixel allows for the study of the equivalent noise charge (ENC) of the chip.
The probability that a pixel responds to a charge can be described by an S-curve function with increasing probability at higher charge.
Noise behaviour can be extracted from the steepness of the curve and the resulting distribution is shown in figure~\ref{fig:lf_gain_noise_irrad} (bottom) before and after irradiation.
The threshold for the two untuned samples are~\SI{7200}{\e} and~\SI{6900}{\e} respectively.
An increase of ENC of about~\SI{150}{\e} can be seen after~\SI{e15}{\neq\per\square\centi\meter}, but the chip can still be operated at a reasonably low threshold to achieve high detection efficiency.

\begin{figure}
    \centering
    \includegraphics[width=.8\columnwidth]{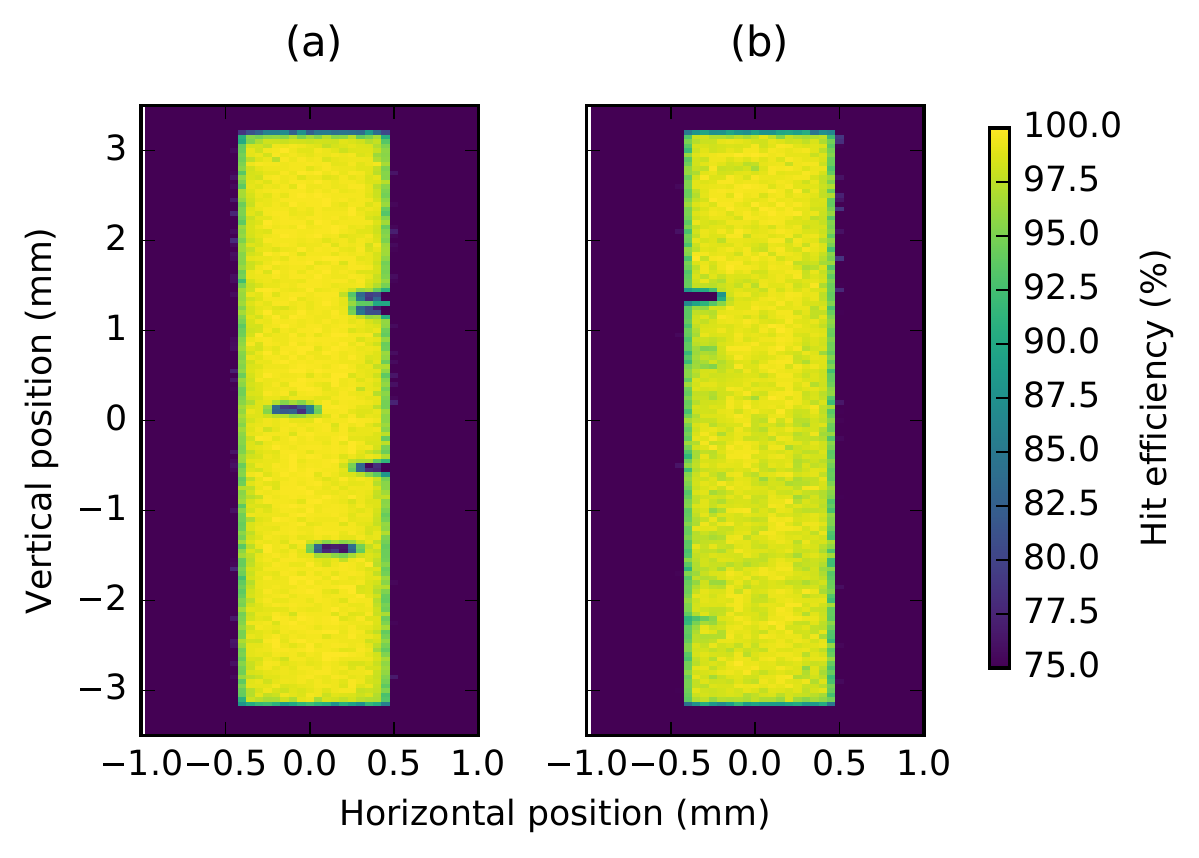}
    \caption{Hit detection efficiency of an (a) unirradiated and (b)~\SI{e15}{\neq\per\square\centi\meter} neutron irradiated LF-Monopix1 chip. From~\cite{lf_paper_toko}.}
    \label{fig:lf_eff}
\end{figure}

Hit detection efficiency for LF-Monopix1 has been reported in~\cite{lf_paper_toko} with average values of~\SI{99.7}{\percent} and~\SI{98.9}{\percent} for the unirradiated and~\SI{e15}{\neq\per\square\centi\meter} neutron irradiated chip at thresholds of~\SI{1800}{\e} and~\SI{1600}{\e}, respectively.
Corresponding efficiency maps are shown in figure~\ref{fig:lf_eff}, where inefficient regions correspond to masked pixels that were excluded from the calculation.

\subsection{Results for TJ-Monopix1}

Similar measurements as for the large collection electrode prototype have been performed with the TJ-Monopix1 chip, including measurements of threshold and noise, hit detection efficiency as well as TID tolerance.
Results have been obtained using a sensor geometry with reduced deep p-well coverage (RDPW) and AC-coupling with additional high voltage applied to the collection electrode to enhance the electrical field in the sensor. 

\subsubsection{Threshold and noise}
\label{sec:tj_thr_enc}

\begin{figure}[htb]
    \centering
    \begin{subfigure}{0.9\columnwidth}
    \centering
        \includegraphics[width=\linewidth]{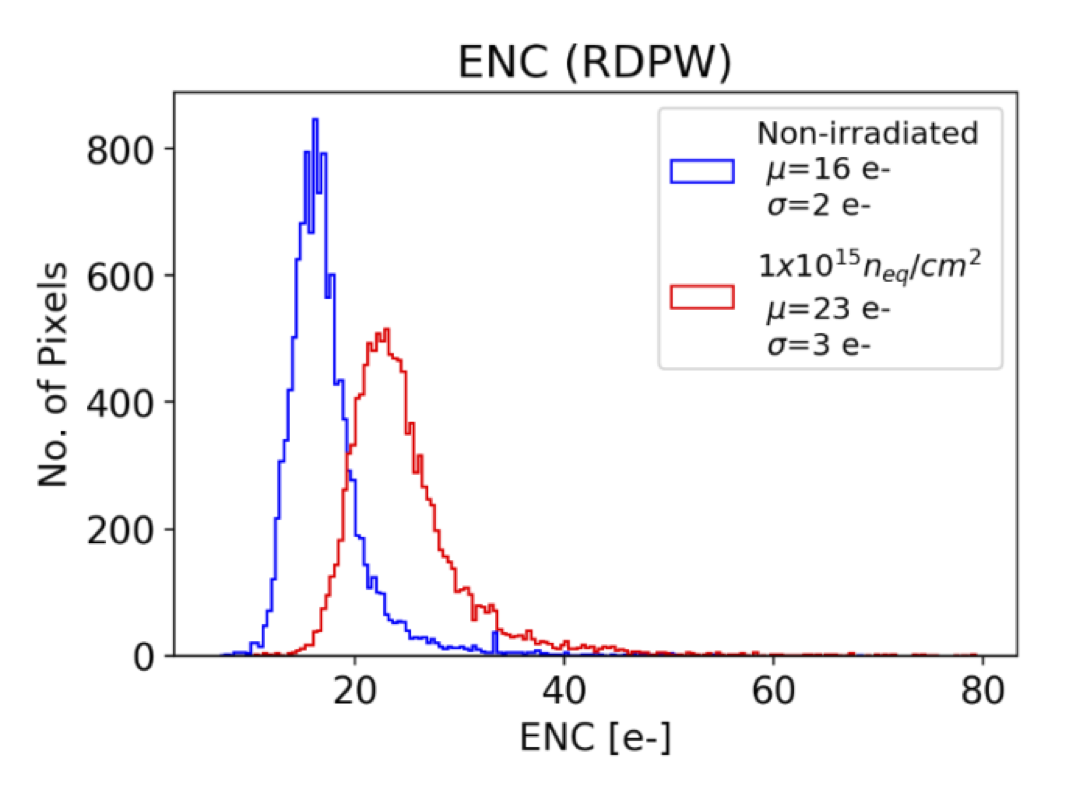}
        \caption{}
        \label{fig:tj_enc}
    \end{subfigure}
    \begin{subfigure}{0.9\columnwidth}
    \centering
        \includegraphics[width=\linewidth]{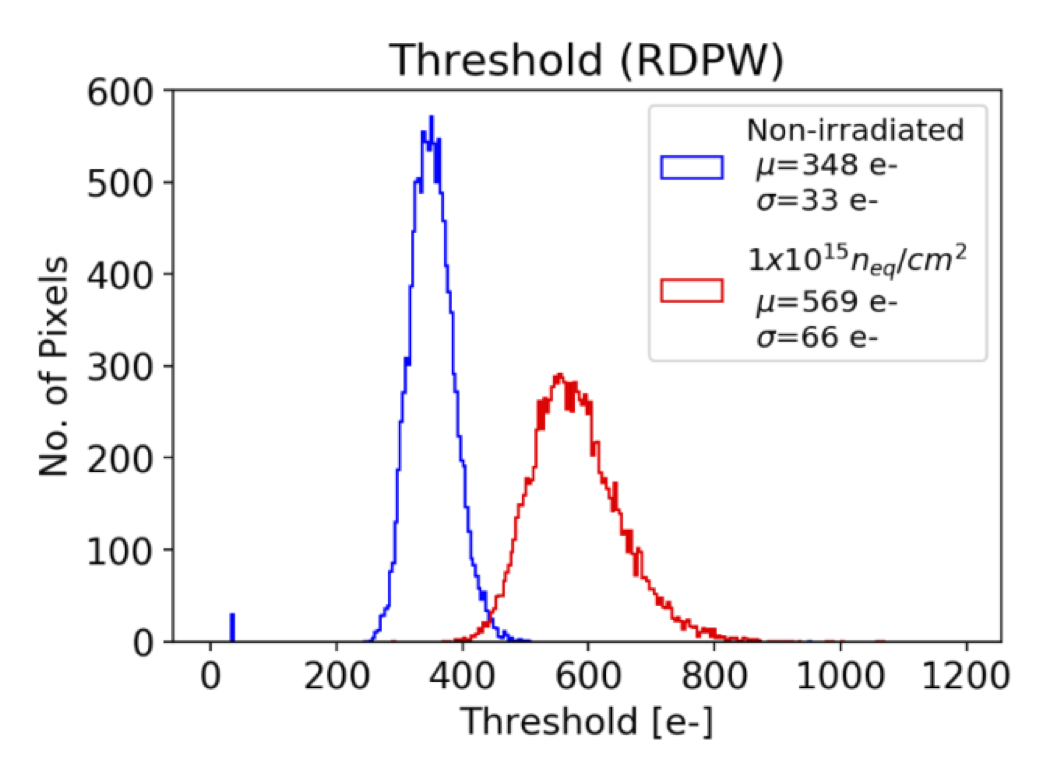}
        \caption{}
        \label{fig:tj_thr}
    \end{subfigure}
    \caption{(a) ENC and (b) threshold distributions of TJ-Monopix1 before and after neutron irradiation. From~\cite{Caicedo_2019}.}
    \label{fig:tj_thr_enc}
\end{figure}

Injecting fixed charge pulses into the pixel electronics allows for the measurement of threshold and noise of the detector.
The mean equivalent noise charge for TJ-Monopix has been reported to be~\SI{16}{\e} before and~\SI{23}{\e} after neutron irradiation~\cite{Caicedo_2019}.
A visual representation is shown in figure~\ref{fig:tj_enc}.
It can be seen that both distributions show an asymmetric tail towards larger ENC values which is an indication for random telegraph signal (RTS) noise~\cite{PhysRevLett.52.228}, which will be improved in a future design.
While the noise can be extracted from the steepness of the function describing the pixel response, the threshold value is defined as the amount of charge, where the probability is~\SI{50}{\percent}.
The resulting distribution for the threshold after converting the result into electrons is shown in figure~\ref{fig:tj_thr}.
The observed threshold increases from~\SI{348}{\e} to~\SI{569}{\e} after neutron irradiation to~\SI{e15}{\neq\per\square\centi\meter} while the spread of the distribution increases by~\SI{100}{\percent} from~\SI{33}{\e} to~\SI{66}{\e}~\cite{Caicedo_2019}.

Due to the increased noise and large threshold dispersion after irra\-diation, operation at low threshold levels is not possible.
This can influence measurements on the hit detection efficiency in case signal charge is lost due to trapping and the remaining detected charge is lower than the threshold.
Changes on the front-end electronics have been studied in a small test chip called MiniMALTA~\cite{Dyndal_2020} that allow for lower operational thresholds as in TJ-Monopix1.

\subsubsection{Hit detection efficiency}

\begin{figure}[htb]
    \centering
    \begin{subfigure}{0.9\columnwidth}
    \centering
        \includegraphics[width=\linewidth]{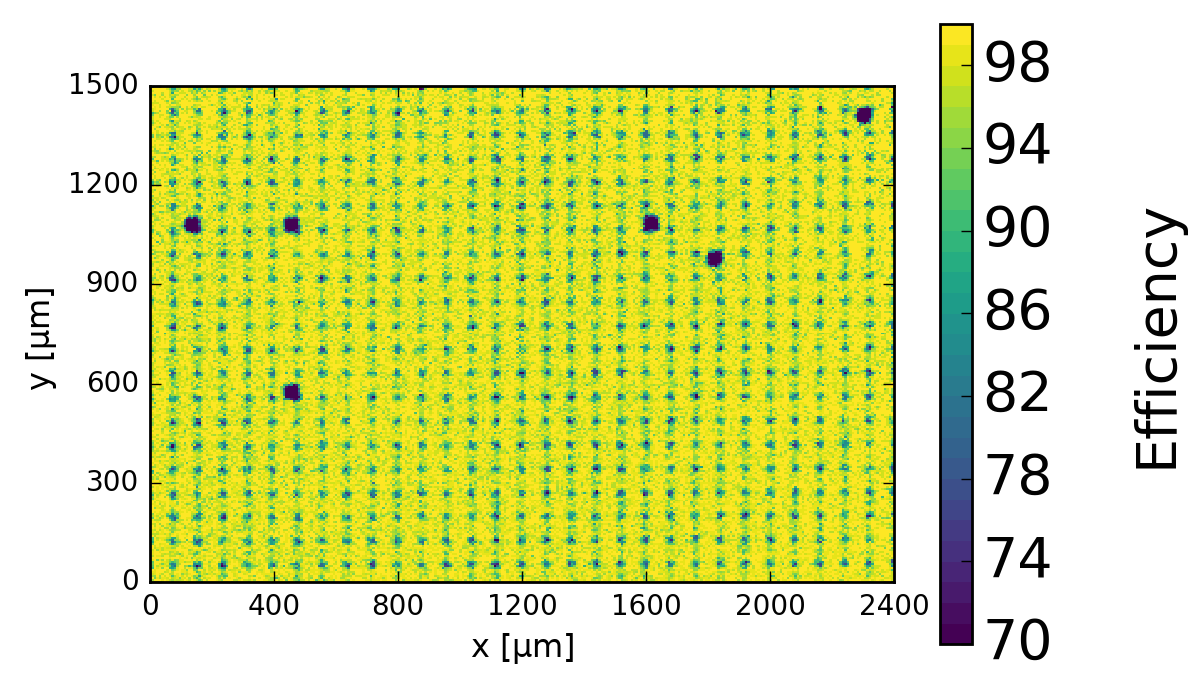}
        \caption{}
        \label{fig:tj_eff_0e15}
    \end{subfigure}
    \begin{subfigure}{0.9\columnwidth}
    \centering
        \includegraphics[width=\linewidth]{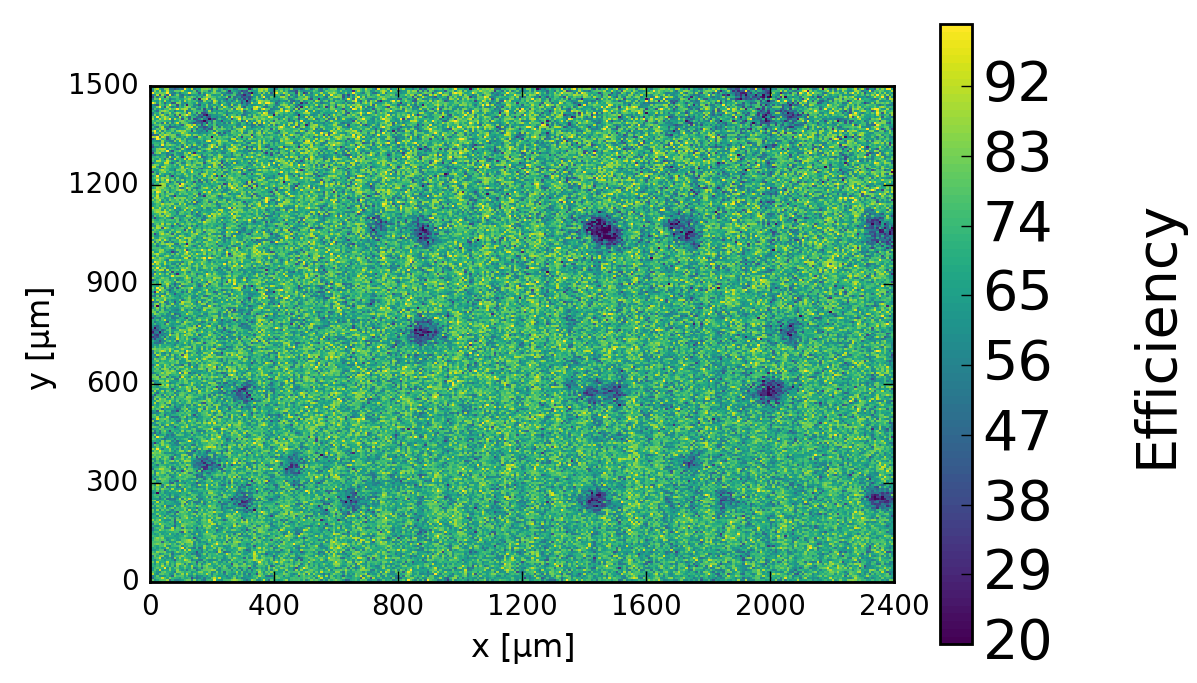}
        \caption{}
        \label{fig:tj_eff_1e15}
    \end{subfigure}
    \caption{Hit detection efficiency of TJ-Monopix1 (RDPW) (a) before and (b) after neutron irradiation to~\SI{e15}{\neq\per\square\centi\meter}. Inefficient regions correspond to masked pixels that were excluded from the efficiency calculation.}
    \label{fig:tj_eff}
\end{figure}

\begin{figure}[htb]
    \centering
    \begin{subfigure}{0.49\columnwidth}
    \centering
        \includegraphics[width=.86\linewidth]{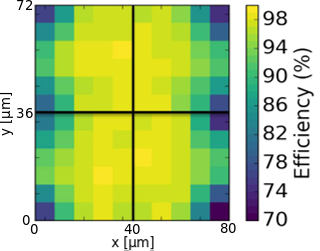}
        \caption{}
        \label{fig:tj_eff_inpix_data}
    \end{subfigure}
    \begin{subfigure}{0.49\columnwidth}
    \centering
        \includegraphics[width=\linewidth]{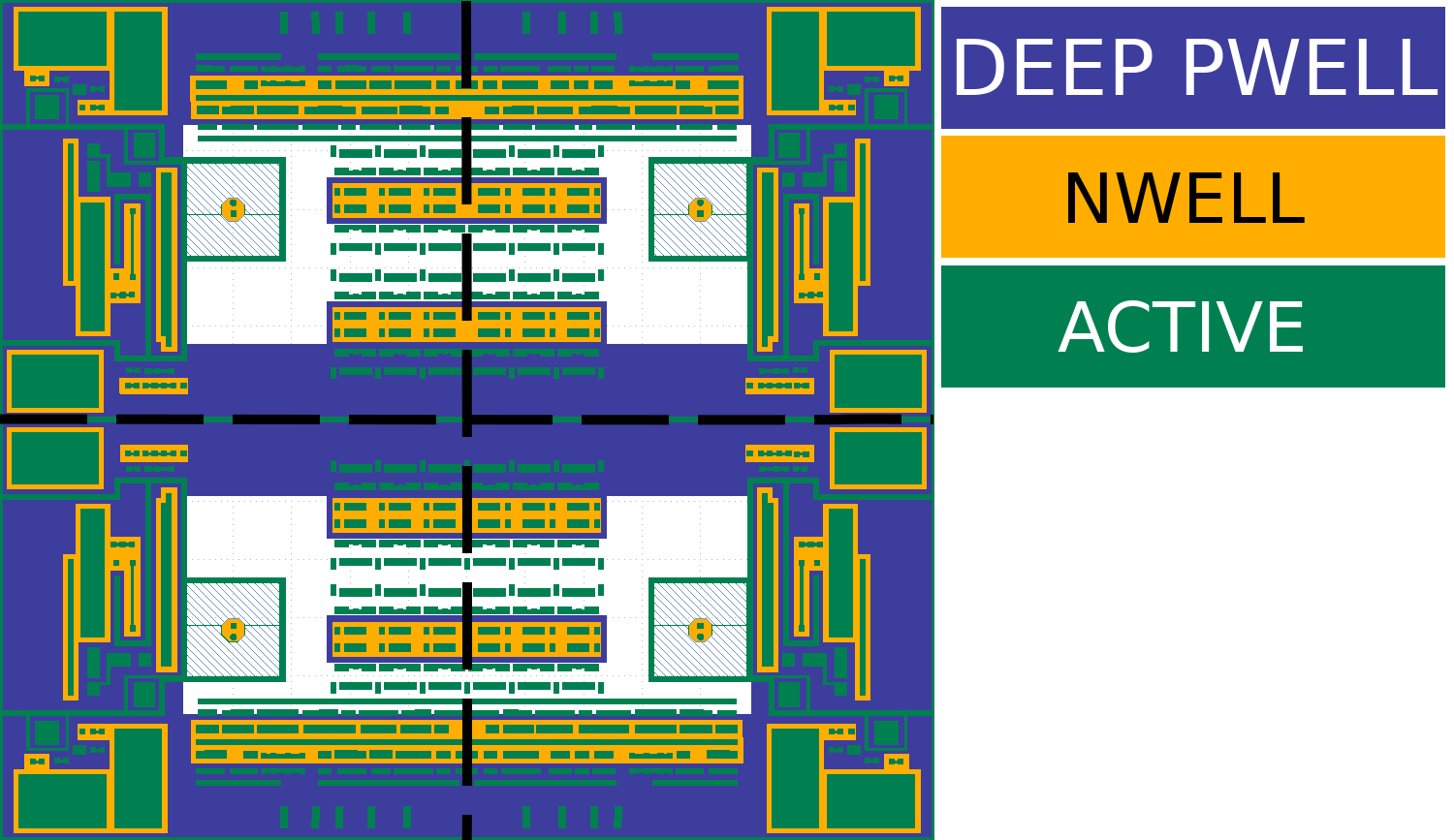}
        \vspace{0cm}
        \caption{}
        \label{fig:tj_eff_inpix_layout}
    \end{subfigure}
    \caption{In-pixel efficiency of an unirradiated 2 x 2 pixel submatrix of TJ-Monopix1 and the layout of such a submatrix showing the relevant areas of deep p-well, n-well and active areas. Inefficient regions correlate with large active areas.}
    \label{fig:tj_eff_inpix}
\end{figure}

The chips have been operated at the threshold settings reported in section~\ref{sec:tj_thr_enc}.
As for LF-Monopix1, noisy pixels were masked to keep the noise occupancy below the ATLAS ITk requirements.
Figure~\ref{fig:tj_eff} shows the resulting efficiency maps for an unirradiated and a neutron irradiated (\SI{e15}{\neq\per\square\centi\meter}) chip with reduced deep p-well coverage.
While the efficiency before irradiation is~\SI{97.1}{\percent} it drops to~\SI{69.4}{\percent} afterwards~\cite{Caicedo_2019}.
In figure~\ref{fig:tj_eff_inpix}, showing the in-pixel efficiency of an unirradiated TJ-Monopix1 chip with reduced deep p-well coverage, a loss of efficiency in the corners and on vertical edges of a 2 x 2  pixel submatrix can be observed.
Studying the design layout of such a submatrix shows a coincidence of inefficient regions with large active areas of the matrix.
Due to a limited spatial resolution of the setup for a cooled irradiated detector during the test beam campaign it was not possible to investigate in-pixel effects for the efficiency loss after irradiation with this detector.
Further studies of in-pixel efficiencies on devices implementing the same sensor geometry~\cite{Cardella_2019} as well as TCAD simulations~\cite{munker_2019} have shown a weak lateral electrical field under the deep p-well areas where charge is lost, especially after irradiation.
There are already two modifications successfully tested to mitigate this effect~\cite{Dyndal_2020}.

\subsubsection{TID tolerance}

\begin{figure}[htb]
    \centering
    \begin{subfigure}{0.9\columnwidth}
    \centering
        \includegraphics[width=\columnwidth]{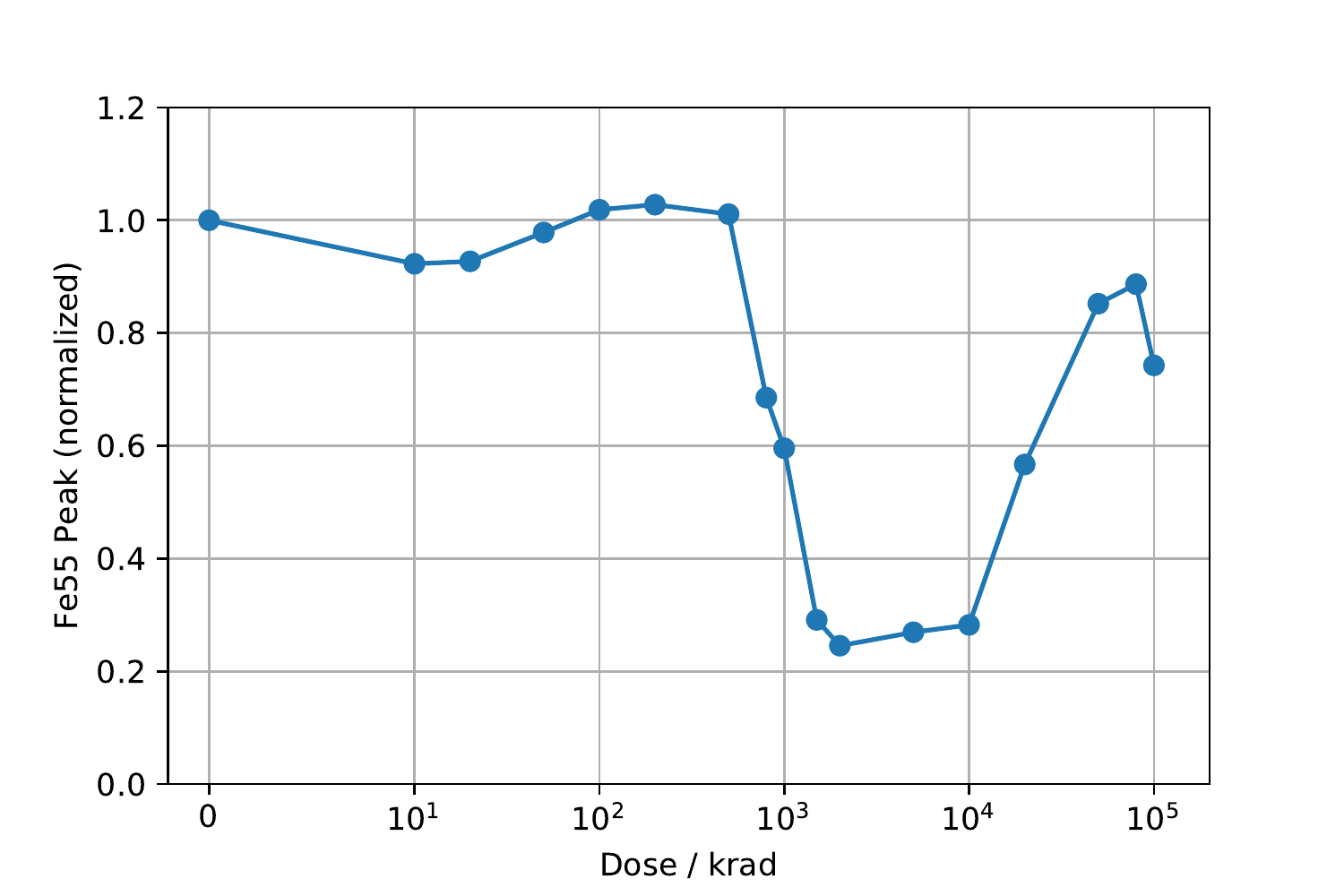}
        \caption{Normalized gain of TJ-Monopix1 depending on TID using a DC coupled frontend with high voltage applied to the p-well and substrate.
        \vspace{1em}}
        \label{fig:tj_tid_gain_pmos}
    \end{subfigure}
    \begin{subfigure}{0.9\columnwidth}
    \centering
        \includegraphics[width=\columnwidth]{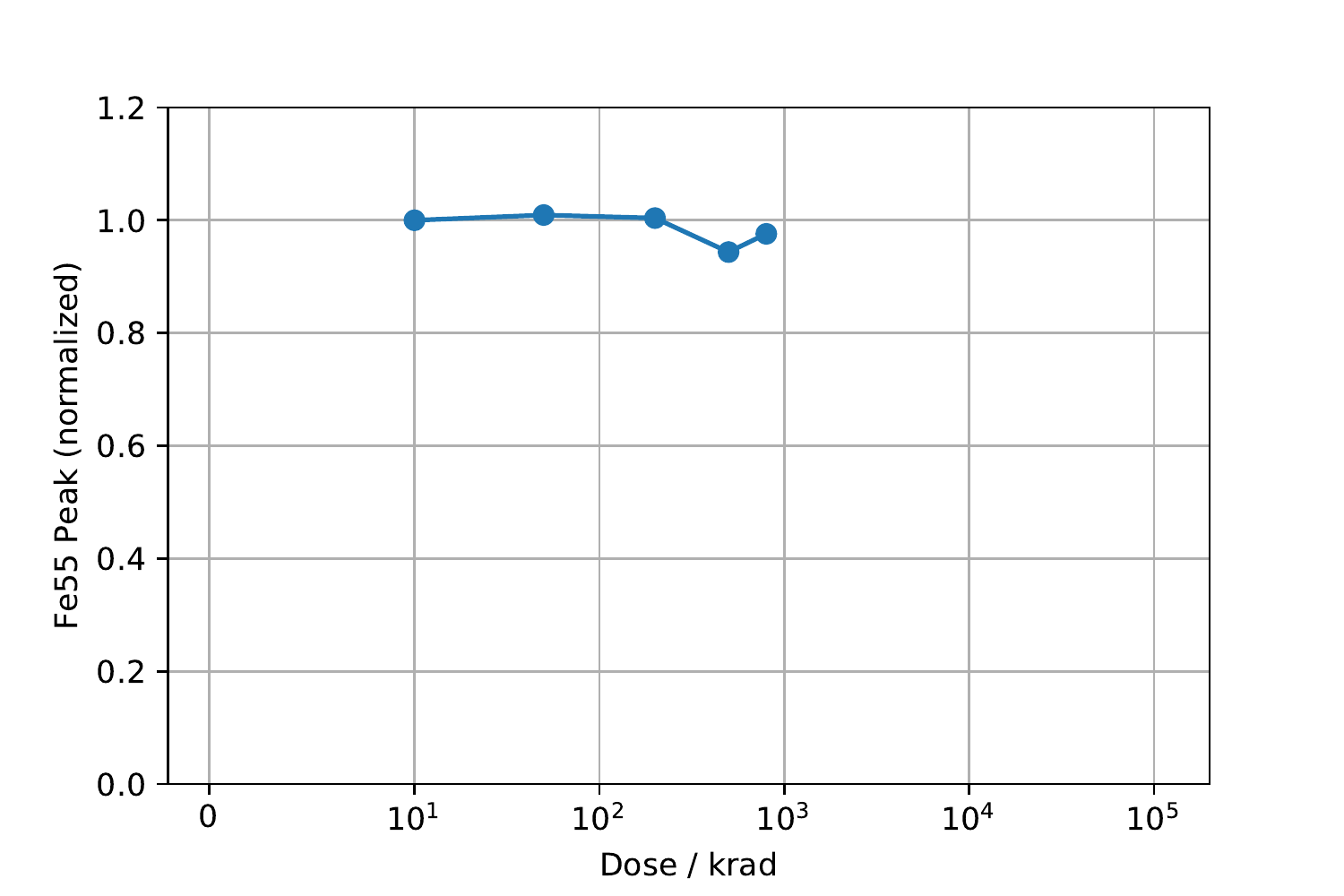}
        \caption{Normalized gain of TJ-Monopix1 depending on TID using an AC coupled front end with additional high voltage on the n-well.}
        \label{fig:tj_tid_gain_hv}
    \end{subfigure}
    \caption{Normalized gain of TJ-Monopix1 versus total ionizing dose measured with an~\ce{^{55}Fe} source. Due to time constraints the sample shown in figure~\ref{fig:tj_tid_gain_hv} was only irradiated to~\SI{0.8}{\mega\rad}.}
    \label{fig:tj_tid_gain}
\end{figure}

The TID tolerance of TJ-Monopix1 has been measured in an X-ray irradiation campaign at Bonn University.
Chips have been irradiated using an X-ray tube at a dose rate of~\SI{0.6}{\mega\rad\per\hour} while being cooled down to approximately~\SI{-3}{\celsius}.
Measurements were conducted directly after every irradiation step.
Figure~\ref{fig:tj_tid_gain_pmos} shows the normalized gain of TJ-Monopix1 in dependence of TID.
After initially stable behavior a large drop of about~\SI{80}{\percent} can be observed between~\SI{0.5}{\mega\rad} and~\SI{10}{\mega\rad}.
A measurement shown in figure~\ref{fig:tj_tid_gain_hv} using a slightly different front-end design with AC coupling and additional high voltage applied to the collection electrode indicates a potentially better radiation tolerance up to the tested~\SI{0.8}{\mega\rad}, not indicating a loss as significant as in figure~\ref{fig:tj_tid_gain_pmos}.
Due to time constraints a measurement to higher doses was not possible and further studies are needed to investigate this observation.

\section{Conclusions}
Large-scale DMAPS prototypes have been characterized for usage at high-rate and high radiation environments like the ATLAS experiment at HL-LHC with promising results.
While LF-Monopix1 shows radiation hardness up to at least~\SI{e15}{\neq\per\square\centi\meter} and consistently high efficiency, the small electrode prototype TJ-Monopix1 shows room for improvements concerning its ra\-diation hardness.
Following the results from both demonstrator chips, these will be improved with larger matrices in both technologies.
The matrix of LF-Monopix2 is increased to~\SI{1x2}{\centi\meter} while reducing the pixel size to~\SI{50x150}{\micro\meter} and optimizing the analog front-end.
It was succesfully submitted in spring~2020.
TJ-Monopix2 will be produced as full size chip of~\SI{2x2}{\centi\meter} with~\SI{33x33}{\micro\meter} pixels and is currently under design.
Process modifications that were tested in a dedicated test chip MiniMALTA will be implemented.
These modifications shows enhanced charge collection, especially after radiation damage and lower operational threshold~\cite{Dyndal_2020}.
The analog front end of TJ-Monopix2 will be improved to reduce noise, increase TID tolerance and allow for lower threshold values as well as adding a threshold tuning on pixel level to reduce threshold dispersion between pixels.

\section*{Acknowledgments}
This project has received funding from the Deutsche Forschungsgemeinschaft DFG, grant WE 976/4-1, 
from the German Ministry BMBF, grant 05H15PDCA9 and the European Union's Horizon 2020 research and innovation programme under grant agreements nos. 675587 and 654168.
\bibliography{mybibfile}

\end{document}